\documentclass[letterpaper,aps,twocolumn,amsmath,amssymb,superscriptaddress]{revtex4}

\usepackage{amssymb}
\usepackage{amsmath}
\usepackage{color}
\usepackage{natbib}

\usepackage[english]{babel}
\usepackage[utf8x]{inputenc}
\usepackage[T1]{fontenc}
\usepackage{wrapfig}

\usepackage{cancel}

\usepackage{graphics}      
\usepackage{graphicx}      
\usepackage{longtable}     
\usepackage{bm}            

\newcommand{\be}{\begin{equation}}
\newcommand{\ee}{\end{equation}}
\newcommand{\bqs}{\begin{equation*}}
\newcommand{\eqs}{\end{equation*}}

\begin{document}
\title{Estimating epidemic arrival times using linear spreading theory}
\author{Lawrence M. Chen}
\affiliation{Department of Mathematics, University of Kansas, Lawrence, KS 66045, USA}
\author{Matt Holzer}
\affiliation {Department of Mathematical Sciences, George Mason University,  Fairfax, VA 22030, USA}
\author{Anne Shapiro}
\affiliation{Department of Mathematics and Statistics, Carleton College, Northfield, MN 55057, USA}
\date{\today}

\begin{abstract} We study the dynamics of a spatially structured model of worldwide epidemics and formulate predictions for arrival times of the disease at any city in the network.  The model is comprised of a system of ordinary differential equations describing  a meta-population SIR compartmental model defined on a network where each node represents a city and edges represent flight paths connecting cities.  Making use of the linear determinacy of the system, we consider spreading speeds and arrival times in the system linearized about the unstable disease free state and compare these to arrival times in the nonlinear system.  Two predictions are presented.  The first is based upon expansion of the heat kernel for the linearized system.  The second assumes that the dominant transmission pathway between any two cities can be approximated by a one dimensional lattice or homogeneous tree and gives a uniform prediction for arrival times independent of specific network features.  We test these predictions on a real network describing worldwide airline traffic.

\end{abstract}

\maketitle


{\bf Modern transportation networks allow for the rapid global spread of disease.  A fundamental property of interest are arrival times.  Given a disease that originates in a particular city or location, how long does it take for that disease to arise in some other city?  In this paper, we develop methods to estimate arrival times based upon the linear determinacy of the system and the theory of traveling fronts.   }

\section{Introduction} 

Mathematical modeling of infectious diseases often involves compartmental models where the total population is divided into groups pertaining to a certain state, i.e susceptible, infected, recovered, etc..  These populations interact and their evolution is described by a system of differential equations, see \cite{kermack32,diekmann00}.   These models can be extended to include spatial structure which often leads to partial differential equation analogs of the local models.  Traditional analysis of the spatial spread of disease has involved the study of traveling fronts; see \cite{hosono95,chen17}.  Ostensibly, the prevalence of long range connections between cities via airline travel would render this approach ineffective.  On the contrary, we demonstrate in this article that when the rate of diffusion is small relative to the reaction rates then the dynamics are still dominated by the formation of traveling fronts only with a different notion of distance between cities to accommodate the long range connections; see  \cite{Brockmann2013,Ianelli2017}.

The model considered here is an SIR (susceptible-infected-recovered) meta-population model; see \cite{rvachev85}.  This model has been employed to great effect to simulate worldwide epidemics; see for example \cite{grais03,hufnagel04,Colizza06,Colizza2007}.  In this model, the world is described by a graph whose nodes correspond to cities and whose edges correspond to flights paths connecting cities.  With local reaction terms describing the disease dynamics at each node and diffusive transport between nodes, the model is a reaction-diffusion equation on a network.  The study of dynamical systems on networks is a burgeoning field and we refer the reader  to \cite{pastor15,barrat08,newman03,porter16} and the references therein.

Our goal in this article is to estimate arrival times of an epidemic spreading through a worldwide airline network.  We make use of the linear determinacy of the system; see \cite{hosono95,chen17}, and apply some tools from linear spreading theory; see \cite{vansaarloos03}, to formulate estimates for the system linearized about the unstable disease free state.  We compare these predictions to observed arrival times in the nonlinear system determined from numerical simulations of the disease spread.

Our methods are most applicable in the case when the rate of diffusion is small compared to the rate of reaction.  In this limit, we are able to obtain explicit predictions for the arrival time of an epidemic in any city as a function of system parameters that is shown to match well with arrival times observed in numerical simulations. Recall that the SIR model under consideration here is nonlinear and high dimensional and it may be surprising that it is possible to explicitly represent such a fundamental characteristic as arrival times.  This is less surprising when the process is viewed through the lens of front propagation into unstable states, where it is often the case that the spreading speeds for the nonlinear system are the same as those in the system linearized about the unstable state.

To summarize, we provide two predictions for arrival times.  Both predictions are obtained from the linearized system near the unstable state.  Representing the solution of the linear system using the heat kernel of the graph Laplacian, we expand the heat kernel and compute estimated arrival times.  The second approach is to approximate the pathway connecting any two cities by a one dimensional lattice or homogeneous tree and then compute the linear spreading speed of solutions propagating along that corridor.  The heat kernel (HK) approach is more accurate as a method of estimation, but requires knowledge of the weights present in the adjacency matrix.  We show that as the diffusion parameter tends to zero, the two estimates agree at first and second order.


Previous authors have also considered disease propagation in complex networks through the perspective of traveling fronts; see for example \cite{hufnagel04,belik11,hindes13}. Our approach is partially motivated by recent work on invasion fronts for the Fisher-KPP equation in random networks; see \cite{hoffman17}.  Previous studies related to estimation of epidemic arrival times include \cite{Gautreau2007,Gautreau2008,Brockmann2013,Ianelli2017}.  These estimates are primarily based upon probabilistic considerations and are related, but distinct, from the methods developed here.  

The paper is organized as follows.  In section~\ref{sec:model}, we review the meta-population SIR model and worldwide airline network that will be the primary object of study.  In section~\ref{sec:arrivaltimes}, we develop two methods to predict arrival times.  In section~\ref{sec:numerics}, we apply our methods to estimate arrival times in numerical simulations of the model.  Finally, we conclude in \ref{sec:conclusion} with a discussion.

\section{Meta-population SIR Model}\label{sec:model}
We study the following meta-population susceptible-infected-reduced (SIR) model described in \cite{Brockmann2013}:
\begin{eqnarray}
\partial_{t}s_{n}&=&-\alpha s_{n}j_{n}+ \gamma\sum_{m\neq n}P_{nm}(s_{m}-s_{n}) \nonumber \\
\partial_{t}j_{n}&=&\alpha s_{n}j_{n}-\beta j_{n}+ \gamma\sum_{m\neq n}P_{nm}(j_{m}-j_{n}) \nonumber \\
\partial_{t}r_{n}&=&\beta j_{n}+ \gamma\sum_{m\neq n}P_{nm}(r_{m}-r_{n}) 
.\label{eq:main} \end{eqnarray}
Here $s_{n}$, $j_{n}$ and $r_n$ are the normalized susceptible, infected and recovered populations of city $n$, respectively.  The parameter $\alpha$ describes the rate of infection and $\beta$ is the mean recovery rate of individuals. The parameter $\gamma$  is the average mobility rate, meaning the percentage of the population of the system traveling at any given time. These parameters govern the spread of the disease. Expressing the dependent variables in vector notation, we see that the disease free state $(\mathbf{s},\mathbf{j},\mathbf{r})=(\mathbf{1},\mathbf{0},\mathbf{0})$ is an equilibrium solution and is unstable so long as $\alpha>\beta$.  As such, if the disease free state is perturbed due to  the introduction of some infected population at one node then the infected population will grow in number and spread into the surrounding network.  

Mathematically, (\ref{eq:main}) constitutes a system of reaction-diffusion equations on a graph $G=(V,E)$.  The nodes correspond to cities and edges represent the presence of air traffic connecting the two cities.  The matrix P is a weighted adjacency matrix with $0\le P_{nm}\le 1$. It is row stochastic, with each $\mathrm{P}_{nm}$ corresponding to the probability of a random walker at node $n$ moving to node $m$.  Due to the row stochasticity of $\mathrm{P}$, observe that total population at each city is preserved and we may recover $r_{n}$ using  $r_{n}=1-s_{n}-j_{n}$.

Since we are interested in epidemics spreading via an airline transportation network, we work with the World Airline Network (WAN) provided by \cite{Guimera2005}.  This network is generated from flights scheduled on approximately 800 airlines for the period November 1, 2000, to October 31, 2001. These flights include both commercial, cargo and smaller air taxi traffic.  While the data is now 16 years old, it can still be considered reflective of the current WAN.   Note that cities are considered as opposed to airports and therefore airports such as O'Hare and Midway are grouped together and considered as the single city Chicago.

The WAN consists of 3618 cities (nodes). There are  14,142 connections (edges). The diameter of the graph is 17, representing the seventeen total flights required to transport from Royal Air Force Brize Norton in Oxfordshire, England, to Wasu, Papua New Guinea.  Properties of this network are studied in \cite{Wu2006,Guimera2005,Crepey2006}, for example.  The network exhibits the small world property and an algebraic degree distribution reminiscent of  random graphs models such as Barbasi-Albert.  Yet at the same time the network retains some spatial structure due to the higher likelihood of connections between cities that are geographically closer.

Since the data only includes connections between airports, we need to specify the matrix $P$ that assigns weights to various connections within the network. Several authors have developed quantitative estimates for these weights based upon city populations and air traffic data; see for example \cite{Brockmann2013}.  We proceed more abstractly and will concentrate on three different types of weights:
\begin{itemize}
\item The \emph{uniform} $\mathbf{P}$ matrix is meant to simulate a simple random walk along the network. As such, in each row every non-zero entry is identical. Here, $\mathbf{P} = \mathbf{D}^{-1}\mathbf{A}$, where $\mathbf{D}$ is the diagonal degree matrix of the network, and $\mathbf{A}$ is the unweighted, undirected adjacency matrix. 
\item The \emph{proportional} $\mathbf{P}$ matrix supposes that a walker may prefer to travel towards nodes with higher connectivity, i.e. local hubs. To implement this, each non-zero entry in each row of $\mathbf{A}$ was assigned a value equal to the column node's degree. The rows were normalized to maintain the matrix's stochastic property.
\item  A \emph{randomized} $\mathbf{P}$ matrix randomly assigns weights to each edge in the network. This was achieved by assigning each non-zero entry in each row of $\mathbf{A}$ a random integer value between 1 and 100, inclusive, and then the rows were normalized. 
\end{itemize}

We now turn our attention to arrival times.

\section{Arrival times: predictions}\label{sec:arrivaltimes}
\begin{figure}[h]
\centering
\includegraphics[width=0.3\textwidth]{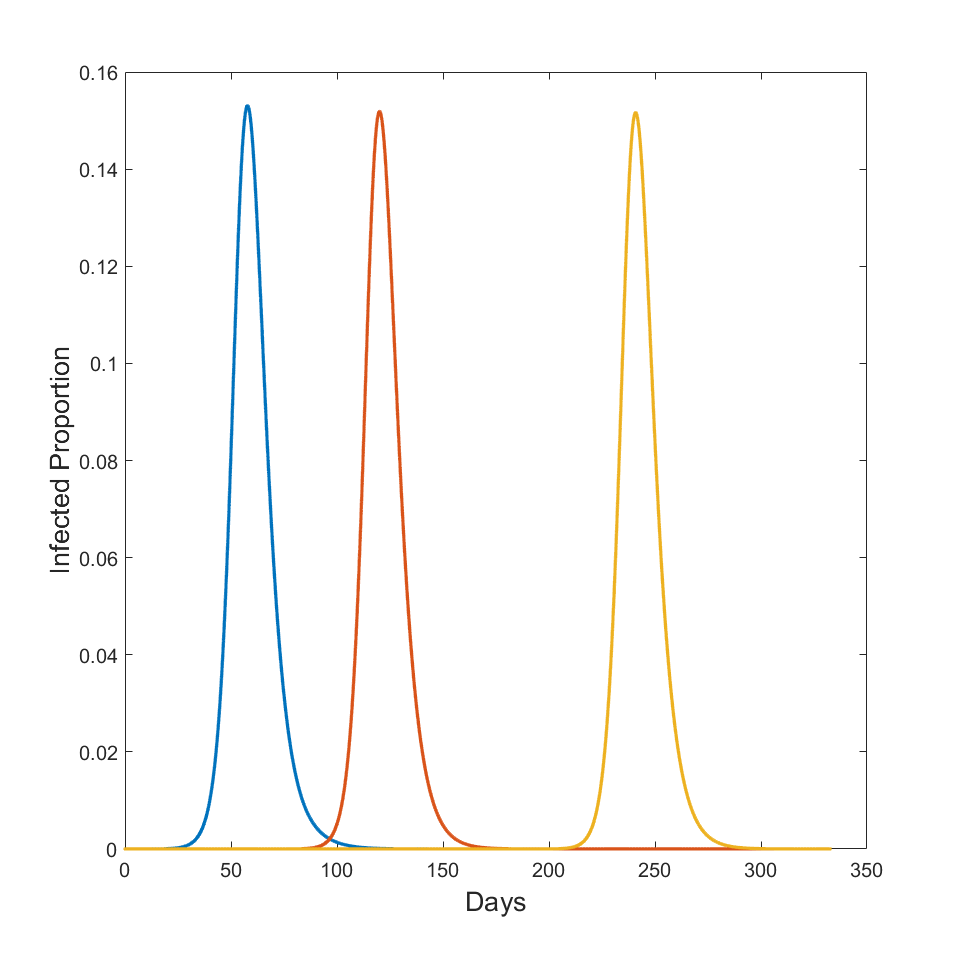} \hfil
\caption{Proportion of infected population in Paris, France (blue), Kodiak, Alaska (Red), and Wasu, Papua New Guinea (Yellow) as a function of time with infection starting in Paris. Simulation was ran using uniform P matrix. Note that in all three cities, the infected population reaches the nearly the same proportion.  Here $\alpha=0.5$, $\beta=0.25$ and $\gamma = 0.001$}\label{fig:solution} 
\end{figure}

Our primary interest in this article is to study how a theoretical epidemic will spread through the network.  To this end, we focus on initial conditions where $j_n(0)=j_0$ for some city $n$ and $j_k(0)=0$ for all other cities. In Figure~\ref{fig:solution}, we plot the infected proportion at three cities (Paris, Kodiak and Wasu) given an epidemic originating in Paris.

We will focus on arrival times, meaning the first time at which a disease is established in a certain city.  Mathematically, for a disease initialized in city $n$ we define the arrival time at city $m$ as 
\be  T_{nm} = \inf \{ \ t\geq 0  \ | \ j_m(t)=\kappa \ \}, \ee
for some threshold $0<\kappa\ll 1$. Note that $T_{nm}$ depends on the specific  cities under consideration,  system parameters, the initial infection proportion $j_0$ and the threshold $\kappa$.  We suppress this functional dependence, but will comment on the dependence on $j_0$ and $\kappa$ later.

Central to our approach is the {\em linear determinacy} of the SIR model (\ref{eq:main}).  Linear determinacy generally applies to reaction-diffusion PDEs models and refers to the property that the speed of invasion fronts in that context can be determined from the linear spreading speed of perturbations in the linearization about the unstable state.  For the spatially extended SIR model, we refer the reader to \cite{hosono95} for studies of invasion fronts in the the PDE case and \cite{chen17} for recent work on a one dimensional lattice.  

To illustrate  linear determinacy in the context of (\ref{eq:main}), we linearize the system about the disease free steady state $(\mathbf{s,j}) = (\mathbf{1,0})$ and obtain 
\begin{eqnarray}
\partial_{t}\mathbf{s}_l&=&-\alpha\mathbf{j}_l +\gamma(\mathrm{P-I})\mathbf{s}_l \nonumber \\ 
\partial_{t}\mathbf{j}_l&=&(\alpha-\beta)\mathbf{j}_l+\gamma(\mathrm{P-I})\mathbf{j}_l. \label{eq:linear} 
\end{eqnarray}
Note that the equation for $\mathbf{j}_l$ decouples and an explicit solution for this linear system can be written as
\be \mathbf{j}_l(t)=j_0e^{\Gamma t}e^{\gamma Pt}v_n,\ee
where we have introduced $\Gamma=\alpha-\beta-\gamma$ and $v_n$ is the standard Euclidean basis vector.  Denote $\mathbf{j}$ as the solution of the nonlinear system (\ref{eq:main}).  By a comparison principle argument we can show that  $\mathbf{j}_l(t)\geq \mathbf{j}(t)$ for all $t>0$, assuming identical initial conditions. To see this, define $Z = \mathbf{j}_l - \mathbf{j}$. The differential equation for the difference $Z(t)$ is,
\be
\frac{dZ}{dt}= -\beta Z+\gamma(\mathrm{P-I})Z+\alpha(\mathbf{j}_l-\mathbf{s}\circ\mathbf{j}),
\ee 
where the product $\mathbf{s}\circ\mathbf{j}$ is the Hadamard product of $\mathbf{s}$ and $\mathbf{j}$ representing component-wise multiplication.  Assume that one component of $Z$ were to become negative.  This is impossible, since when that component was zero we would necessarily have that $\frac{dZ}{dt}>0$ due to the positivity of $Z$ at nearby nodes and the component wise bound $\mathbf{s}\leq 1$.  

This leads to the important observation that the arrival times for the linear problem always provide  lower bounds on the arrival times in the nonlinear problem. If the initial condition is small then it will turn out that this lower estimate is in fact close to the actual arrival times in the nonlinear system.  Mathematically, we reason this to be due to the fact that when $j_0\ll 1 $ then the system is well approximated by the linear system for a sufficiently long period of time.  From the point of view of the application, this condition translates to having precise information on the time of origination of the disease.

We therefore concern ourselves with determining the arrival times for the linear system, i.e we wish to estimate
\be \inf \{t\geq 0 \ | \  j_0v_m^Te^{\Gamma t}e^{\gamma Pt}v_n=\kappa \}. \label{eq:hkat} \ee
We now proceed to derive estimates  for arrival times based upon the solution of  the linear problem. We will return in Section~\ref{sec:numerics} to compare these estimates to arrival times in the full nonlinear system.

\subsection{Prediction via Heat Kernel Expansion}

We now specify a prediction for arrival times based upon the linear dynamics that we expect to be accurate when the rate of diffusion is small relative to the reaction rates.  The method is based upon expansion of the heat kernel in (\ref{eq:hkat}).  

Consider the full linear evolution of the infected population in (\ref{eq:linear}),
\be j_m(t)=j_0v_m^Te^{\Gamma t}e^{\gamma \mathrm{P}t}v_n,\ee
where again recall that we have assumed that the infection originates in city $n$ with initial infected proportion $j_0$.  We expand the matrix exponential $e^{\gamma \mathrm{P}t}$ as a series
\be v_m^T e^{\gamma \mathrm{P}t}v_n=\sum_{k=0}^\infty   
 \frac{\gamma^k t^k}{k!} v_m^T\mathrm{P}^k v_n. \label{eq:sum} \ee
Let $\rho_k = v_m^T\mathrm{P}^k v_n$.  Note that $\rho_k$ describes the probability of a random walker starting at city $m$  being located in city  $n$ after $k$ steps.  Suppose that the minimal number of flights required to travel from city $n$ to city $m$ is $d$. Then, in the expansion 
\be 
	j_m(t)= j_0e^{\Gamma t}\left[\rho_0 + \gamma\rho_1t + \dots + \frac{\gamma^d \rho_d t^d}{d!} + \dots\right],\label{eq:jm}
\ee
all terms before the $d$th term must be zero. 

To formulate an arrival time prediction based upon the heat kernel expansion, we assume that the leading order term in the expansion dominates for small $\gamma$.  That is, the dominant contribution to the disease spreading from node $n$ to $m$ is along the shortest path(s) of length $d$.

By solving, 
\be j_0e^{\Gamma t} \frac{\gamma^d \rho_d t^d}{d!}=\kappa, \ee
we obtain the heat kernel estimate for the arrival time
\be 
	T_{nm}^{HK} = \frac{d}{\alpha-\beta-\gamma}W\left(\frac{(d !)^{1/d}}{d}\frac{\alpha-\beta-\gamma}{\gamma (\rho_d)^{1/d}}\left(\frac{\kappa}{j_0}\right)^{1/d}\right), \label{eq:THK}
\ee
where $W$ is the Lambert-W function.

Validity of the heat kernel prediction depends on the leading order term in (\ref{eq:sum}) dominating.  There are two general ways in which this assumption can fail. The first involves cases where $\rho_d\ll \rho_{d+1}$ for some pair of cities, suggesting that contribution from  the second term in the heat kernel expansion can not be ignored.  A second case occurs for larger values of $\gamma$, where accurately estimating the sum (\ref{eq:sum}) requires multiple terms.  Therefore, we expect the heat kernel estimate to closely approximate the linear dynamics in the asymptotic limit as $\gamma \to 0$.  If $\gamma$ is no longer small, then we see that the shortest paths need not be the most dominant contributor to disease spread through the network. This deficiency can be remedied by including more terms in the the heat kernel. For example, we could consider
\begin{align}
 e^{\Gamma t}\left[\frac{\gamma^d\rho_dt^d}{d!}\left(1 + \frac{\gamma t}{(d+1)}\frac{\rho_{d+1}}{\rho_d} +\dots\right)\right] = \frac{\kappa}{j_0}.
\end{align}
Instead of neglecting the terms after the $d$th term, the estimate can be refined to include as many terms as needed.  

In Section~\ref{sec:numerics}, we illustrate the accuracy of the heat kernel estimate in the context of the WAN.  Before doing so, we comment on the relationship between this prediction and the theory of invasion fronts and compare this prediction to other estimates in the literature.

\subsection{Linear Spreading Speed}
We now derive a estimate for arrival times based upon the calculation of the linear spreading speed in certain lattice dynamical systems.  
We have two main goals here.  The first is to establish a theoretical connection between the heat kernel estimate provided in (\ref{eq:THK}) and the theory of traveling invasion fronts.  Secondly, we note that the heat kernel estimate requires explicit knowledge of the weights in the network through the quantities $\rho_d$.  In situations where the network topology is known, but the particular weights are not then we desire a method to estimate average arrival times for these networks.

The key assumption here is that the shortest path(s) -- in terms of flights required -- connecting the cities $m$ and $n$ constitute the dominant transmission pathway of the disease between those cities.  This corridor is then assumed to be well approximated by a one dimensional lattice or a homogeneous tree.  For the one dimensional lattice this assumption translates to one where the loss of infected population to the remaining network is either negligible or balanced by transport in the opposite direction.  For the homogeneous tree, the assumption is that the loss of infected population to the rest of the network can be suitably approximated by a tree.  

Under this assumption the dynamics of (\ref{eq:linear}) along the shortest path are approximated by the following  linear, lattice dynamical system
\begin{align}
\partial_t j_{n}=&(\alpha-\beta)j_n+\frac{\gamma}{k+1}[j_{n-1}-(k+1)j_{n}+kj_{n+1}]. \label{eq:lattice}
\end{align}
Here $k=1$ for a one dimensional lattice and $k>1$ for a homogeneous tree with branching factor $k$.  To determine the linear spreading speed of solutions of (\ref{eq:lattice}), we first substitute the ansatz $j_n(t)=e^{\lambda t-\nu n}$, from which the dispersion relation,
\begin{align}
\lambda&=(\alpha-\beta)+\frac{\gamma}{k+1}[e^{\nu}-(k+1)+ke^{-\nu}]
\end{align}
relates solutions with spatial decay rate $\nu$ to their corresponding temporal growth rate $\lambda$.  With $\lambda(\nu)$ in hand, the envelope speed of $e^{-\nu n}$ is
\be s_{env}(\nu)=\frac{\lambda(\nu)}{\nu}.\label{eq:senv} \ee
The envelope speed gives the speed at which an exponentially decaying solution $e^{-\nu n}$ spreads in the linear system.  Minimizing over all such $\nu$ we obtain the {\em linear spreading speed},
\be s_{lin}=\min_{\nu>0} s_{env}(\nu).\ee
Differentiating (\ref{eq:senv}) with respect to $\nu$, we see that critical points can be found by simultaneously solving the system of equations 
\be s=\frac{\lambda(\nu)}{\nu}, \quad s=\partial_\nu \lambda(\nu). \label{eq:PDR} \ee
Roots of this system of equations are known as pinched double roots; see \cite{vansaarloos03,holzer14}.  Setting $\frac{\lambda(\nu)}{\nu}=\partial_\nu \lambda(\nu)$, we obtain the following equation,
\begin{align}
0&=(\alpha-\beta)+\frac{\gamma}{k+1}[(1-\nu)e^\nu -(k+1)+(1+\nu)ke^{-\nu}]. \label{eq:PDR}
\end{align}
We write $\nu_{lin}(\gamma)$ as the solution of (\ref{eq:PDR}), from which 
\be s_{lin}(\gamma)=\frac{\gamma}{k+1}\left( e^{v_{lin}(\gamma)}-ke^{-\nu_{lin}(\gamma)}\right). \label{eq:slin} \ee
In general, we do not have analytical formulas for the linear spreading speed and $s_{lin}(\gamma)$ must be computed numerically.  The linear spreading speed  (LSS) approximation provides one spreading speed throughout the entire network for any given set of initial parameters. The quantity $\frac{1}{s_{lin}(\gamma)}$ then provides an estimate for the  inter-city arrival times in the network. Multiplying by the minimal number of flights, $d$, required to traverse the network from city $n$ to city $m$,  we can obtain a  prediction for the arrival times, which we denote
\be T_{nm}^{LSS}=\frac{d}{s_{lin}(\gamma)} \ee

The arrival time estimate $T_{nm}^{LSS}$ does not incorporate any specific information concerning the topology or weights of the  transportation network.  This is both a strength and a weakness.   On the positive side,  less information is required in order  to formulate a prediction for arrival times.  Of course, on the other hand, we expect the accuracy of this prediction to suffer due to the limited details of the network incorporated.

\subsection{Comparison of predictions}

We now compare the two estimates  $T_{nm}^{HK}$ and $T_{nm}^{LSS}$.  For simplicity, we set $j_0=\kappa$; see the discussion in the second paragraph of Section~\ref{sec:numerics}.

Recall that we are interested in the limit $\gamma\to 0$ and focus on that case here.  We first obtain expansions for $T_{nm}^{LSS}$ in this limit.  Recall Equation (\ref{eq:PDR}).  Assuming that $\alpha-\beta$ is positive and independent of $\gamma$, we then require that $\nu\to \infty$ to balance the asymptotically small $\gamma$ terms.  In doing so, we must solve
\be \alpha-\beta=\frac{\gamma}{k+1}(\nu-1)e^\nu. \ee
This yields an estimate for the linearly selected decay rate
\be \nu_{lin}\approx W\left(\frac{(\alpha-\beta)(k+1)}{e\gamma}\right)+1,\ee
where $W$ is the Lambert-W function.  Since the arrival time is approximately $\frac{d}{s_{lin}}$, we see from  (\ref{eq:slin}) that 
\begin{align} T_{nm}^{LSS} \approx \frac{d}{\alpha-\beta}W\left(\frac{(\alpha-\beta)}{e\gamma}(k+1)\right)\label{eq:TLSS}
\end{align}

Therefore, both $T_{nm}^{HK}$ and $T_{nm}^{LSS}$ can be expressed in terms of Lambert-W  functions.  Asymptotic expansions of both expressions are readily available using properties of the Lambert-W function.  We obtain,
\begin{eqnarray} T_{nm}^{LSS} &=& \frac{-d}{\alpha-\beta}\log\gamma-\frac{d}{\alpha-\beta}\log(-\log\gamma) \nonumber \\
&+&\frac{d}{\alpha-\beta}\log\left(\frac{(\alpha-\beta)(k+1)}{e}\right)+o(1), \end{eqnarray}
and
\begin{eqnarray} T_{nm}^{HK} &=& \frac{-d}{\alpha-\beta}\log\gamma-\frac{d}{\alpha-\beta}\log(-\log\gamma) \nonumber \\
&-&\frac{d}{\alpha-\beta}\log\frac{d}{\alpha-\beta}-\frac{1}{\alpha-\beta}\log\frac{\rho_d}{d!}+o(1).\end{eqnarray}
As $\gamma\to 0$ arrival times tend to infinity.  Both arrival time estimates agree to first and second order in their expansions.  They differ at $\mathcal{O}(1)$.  At this juncture, we compare our arrival time estimates to those of \cite{Gautreau2008}.   In fact, three different arrival time estimates are provided in \cite{Gautreau2008}: referred to as front, Gumbel and deterministic. The deterministic formulation  involves the Lambert-W function and is the most related to the heat-kernel and linear spreading speed estimates provided here.  In the limit as $\gamma\to 0$, the deterministic estimate of \cite{Gautreau2008}, the  heat kernel and linear spreading speed estimates are equivalent to $\mathcal{O}(\log(-\log(\gamma)))$ but differ at $\mathcal{O}(1)$.  The Gumbel estimate  is equivalent to $T_{nm}^{LSS}$ and $T_{nm}^{HK}$ to leading order, but has no $\log(-\log(\gamma))$ correction.  

Note that the dependence of $T_{nm}^{LSS}$ on the branching factor $k$ does not appear until $\mathcal{O}(1)$.  We comment on the choice of this value.  Comparing expressions (\ref{eq:TLSS}) and (\ref{eq:THK}) directly, we  could match the arguments of the Lambert-W functions
\be \frac{\alpha-\beta}{e\gamma}(k+1)=\frac{(d !)^{1/d}}{d}\frac{\alpha-\beta}{\gamma (\rho_d)^{1/d}}.\ee
Note that for large $d$
\be \frac{(d!)^{1/d}}{d} \approx e^{-1}.\ee
This suggests that $k$ should be chosen according to 
\be k+1=\frac{1}{(\rho_d)^{1/d}}.\ee 
Thus, if there was a single shortest path connecting two nodes, the tree with degree $k+1$ given by the geometric mean of the weights connecting the two cities is a likely choice.  Since typical shortest paths pass through high degree hubs, this would suggest large values of $k$ in general.  That being said, it is rarely the case that there is a single shortest path connecting two cities.  In the numerical simulations in section~\ref{sec:numerics}, we typically take $k$ to be one less than the average degree in the network.  For the WAN from \cite{Guimera2005}, this is $k=6.871$.

\section{Arrival times: predictions versus numerical observations}~\label{sec:numerics}

We now apply our predictions to estimate arrival times for the nonlinear system (\ref{eq:main}).  For the purposes of simulations, we fix the parameters $\alpha=0.5$ and $\beta=0.25$.  We vary $\gamma$ between $0.001$ and $0.1$ and note that this range  includes the estimated range for airline traffic  derived in \cite{Brockmann2013}.  Numerical simulations are performed using explicit Euler with typical timestep $\Delta t=0.01$.  

A remark regarding the initial condition $j_0$ and threshold $\kappa$ is in order.  As a general rule, the smaller these quantities are the better the estimates are.  In simulations, we have typically set $j_0=\kappa$ and used values between $0.01$ (arrival when one percent of the population is infected) and $10^{-6}$ (arrival when approximately one individual is infected).  This range of values also give satisfactory results even if $j_0\neq \kappa$.  Too large of values for $j_0$ can be problematic since the assumption is that the system is well approximated by its linearization, at least on the portion of the network where the interesting dynamics are occurring, i.e. ahead of the front interface.  If one insists on larger values of $j_0$, then there is typically some initial transient before the estimates are valid.  For the linear spreading speed estimate this may not be an issue, however in the heat kernel estimate this can mean that the estimates are off by some fixed constant.

\subsection{Uniform $\mathrm{P}$: Heat Kernel Estimate}

We first consider the case of $\mathrm{P}=\mathrm{D}^{-1}\mathrm{A}$ describing a random walk on the network where weights at each node are chosen evenly. 

\begin{figure}[h]
\centering
\includegraphics[width=0.4\textwidth]{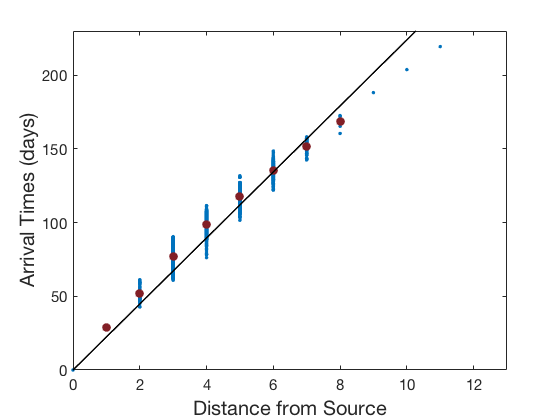}
\includegraphics[width=0.4\textwidth]{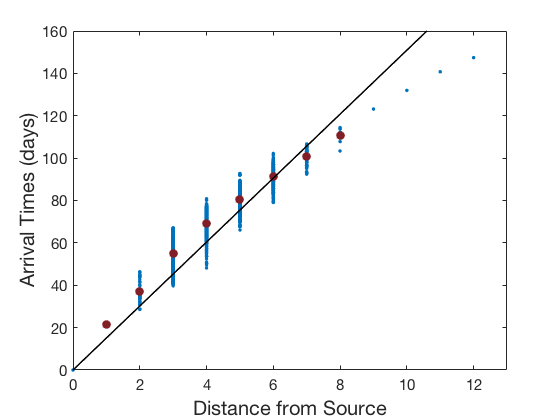} 
\hfil
\caption{Arrival times compared to distance from infection source node with $\gamma=0.001$ (upper) and $\gamma=0.01$ (lower) and infection originating in Kodiak, Alaska. Red dots represents mean arrival time at each difference. The ultimate 4 nodes represent cities in Papua New Guinea and were not included in analysis. For $\gamma=0.001$, the regressed slope gives arrival times of 20.779 between cities while the linear spreading speed predicts arrival times of 19.961 days.  For $\gamma=0.01$, the regressed slope gives arrival times of 13.435  between cities while the linear spreading speed predicts arrival times of 12.920  days. In both cases $\alpha=0.5$ and $\beta=0.25$. }\label{fig:kodiakLSS} 
\end{figure}

\begin{figure}[h]
\centering
\includegraphics[width=0.4\textwidth]{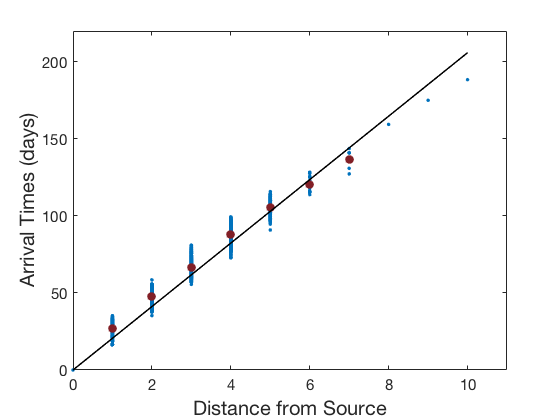} \hfil
\includegraphics[width=0.4\textwidth]{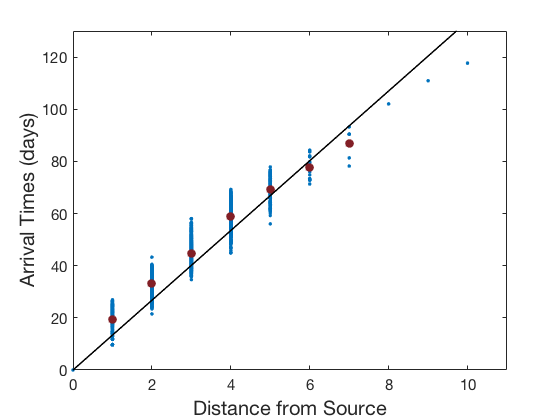} \hfil
\caption{Arrival times compared to distance from infection source node with $\gamma=0.001$ (upper) and $\gamma=0.01$ (lower) with an infection originating in Paris, France. Red dots represents mean arrival time at each difference. The ultimate 3 nodes represent cities in Papua New Guinea and were not included in analysis. For $\gamma=0.001$, the regressed slope gives arrival times of 19.223 between cities while the linear spreading speed predicts arrival times of 19.961 days.  For $\gamma=0.01$, the regressed slope gives arrival times of 11.950 between cities while the linear spreading speed predicts arrival times of 12.920  days. In both cases $\alpha=0.5$ and $\beta=0.25$.}\label{fig:parisLSS} 
\end{figure}

To focus our discussion, we will consider infections starting in only two  locales: Paris, France and Kodiak, USA.  We select these nodes since Paris is the city with the largest number of connections and Kodiak is a periphery node with only one connection to Anchorage. Arrival times as a function of distance from the initial node are shown in Figure~\ref{fig:kodiakLSS} for Kodiak and in Figure~\ref{fig:parisLSS} for Paris.  Note the general linear relationship exhibited by the data with nodes further from the source taking longer for the infection to arrive at these nodes than closer nodes.  Simultaneously, we note that there remains significant deviations in the arrival times of the infection even among nodes that are a fixed distance from the source node. We will compare these observed arrival times to those predicted by the heat kernel and linear spreading speed methods.  

\begin{figure}[ht]
\centering
\includegraphics[width=0.4\textwidth]{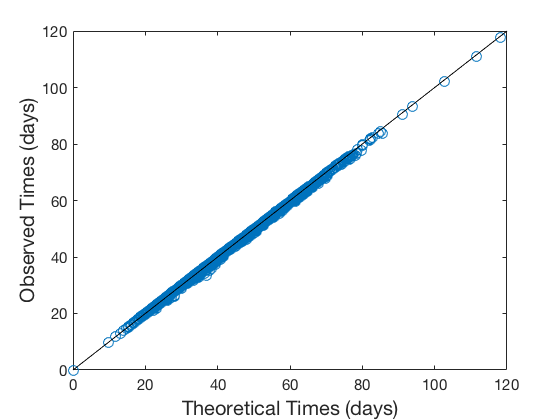} \hfil
\caption{Simulated arrival times compared to one-term heat kernel approximation using uniform P matrix for infection originating in Paris, France with $\gamma=0.01$ plotted against the line $y=x$. }\label{fig:parisHK01} 
\end{figure}

\begin{figure}[h]
\centering
\includegraphics[width=0.4\textwidth]{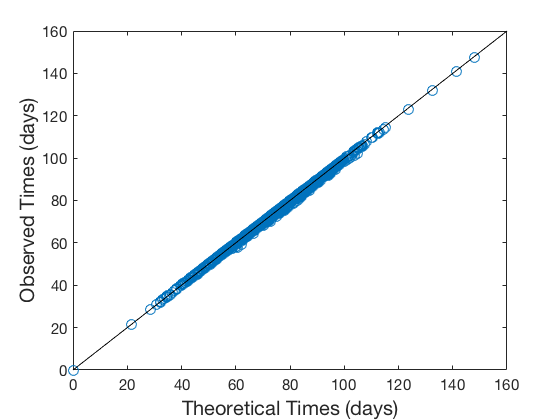} \hfil
\caption{Simulated arrival times compared to one-term heat kernel approximation using uniform P matrix for infection originating in Kodiak, Alaska with $\gamma=0.01$ plotted against the line $y=x$. }\label{fig:kodiakHK01} 
\end{figure}

For sufficiently small values of $\gamma$, we find that  arrival times for (\ref{eq:main}) observed in numerical simulations agree well with predictions based upon the heat kernel expansion.  In Figure~\ref{fig:parisHK01}, predicted arrival times versus observed arrival times for an infection originating in Paris are plotted and we find excellent agreement between the two with mean absolute errors of approximately $0.40$ days.  Correspondence is even better for smaller values of $\gamma$.  Similar results hold for infections originating in Kodiak, see Figure~\ref{fig:kodiakHK01}.

For $\gamma$ values larger than $0.01$, we find that the heat kernel approximation also exhibits errors.  We work with a specific example of an infection originating in Kodiak and arriving at Yellowknife, Canada.  The shortest path between these two cities consists of four flights and  involves first flying to Anchorage, followed by one of five hubs (Chicago, Los Angeles, Minneapolis-St. Paul, Seattle, Toronto), followed by flights to Edmonton and then Yellowknife.  For asymptotically small values of $\gamma$ this is the dominant transmission path.  However, for larger values of $\gamma$ there are longer paths with non-negligible contribution to the arrival times.  For example, one can fly from Anchorage to Juneau to Whitehorse to Fort Simpson to Yellowknife. This route requires five flights, but the number of flights emanating from these cities is very low compared to the shorter path.  This leads to $\frac{\rho_5}{\rho_4}\approx 8$ for this example, which translates to saying that it is eight times more likely for a random walker to traverse the network and arrive in Kodiak in five steps as opposed to four.  Therefore, as $\gamma$ increases the most probable path is not necessarily the shortest.

\begin{figure}[ht]
\centering
\includegraphics[width=0.44\textwidth]{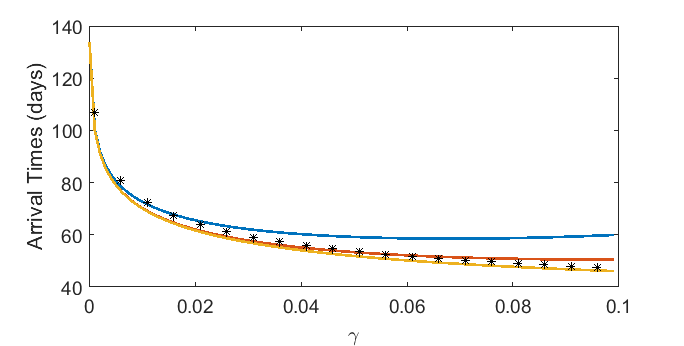} \hfil
\caption{Arrival times in Yellowknife as a function of $\gamma$ for an infection starting in Kodiak (asterisks).  In blue is the one term heat kernel approximation, orange is the two term heat kernel and the fifteen term heat kernel is in yellow.   }\label{fig:kodiakHK2} 
\end{figure}

A second issue is that when $\gamma$ is not asymptotically small, one may require more terms in the heat kernel expansion to get a reasonable approximation.  For example, supposing that the arrival time scales with $\log\gamma$, then the terms $(\gamma t)^k$ do not converge rapidly to zero and more than one term is likely required to get an accurate approximation.  In Figure~\ref{fig:kodiakHK2}, we plot arrival times in Yellowknife for an infection originating in Kodiak.  The one term heat kernel exhibits errors due to the mechanism just described, but the two term heat kernel closely matches the full heat kernel estimate over a large range of $\gamma$ values.  

\begin{figure}[ht]
\centering
\includegraphics[width=0.4\textwidth]{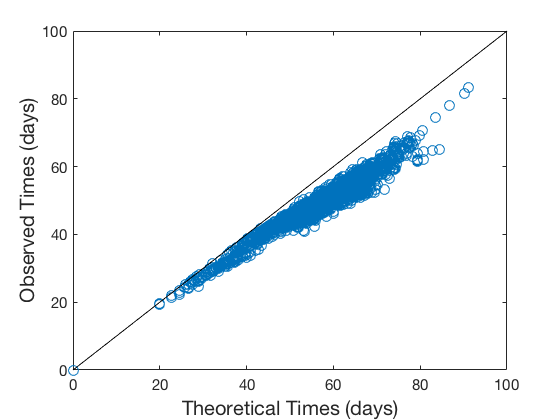} \hfil
\caption{Simulated arrival times compared to one-term heat kernel approximation using uniform P matrix for infection originating in Kodiak, Alaska with $\gamma=0.1$ plotted against the line $y=x$. }\label{fig:kodiakHK1} 
\end{figure}

\begin{figure}[ht]
\centering
\includegraphics[width=0.4\textwidth]{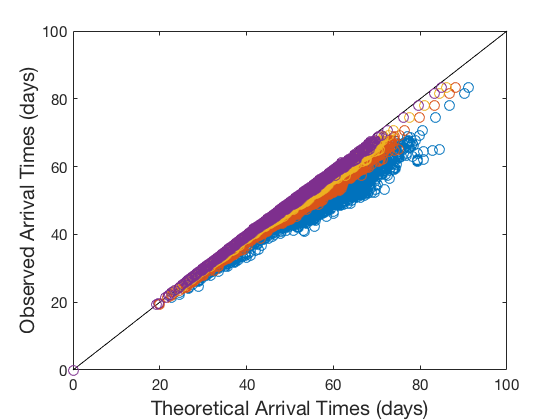} \hfil
\caption{Simulated arrival times compared to one, two, three, and four-term heat kernel approximations using uniform P matrix for infection originating in Kodiak, Alaska with $\gamma=0.1$ plotted against the line $y=x$. One-term heat kernel is plotted in blue, two term is plotted in orange, three term is plotted in yellow, and four term is plotted in purple As more terms are added to the heat kernel approximation, it grows closer to the line.}\label{fig:kodiakHKMT} 
\end{figure}

In Figure~\ref{fig:kodiakHK1}, one observes the errors that are present in the one term heat kernel expansion for an epidemic initialized in Kodiak.  Here $\gamma=0.1$ is unrealistically large for the application at hand, but we include it for illustrative purposes. We see that including more terms in the heat kernel expansion leads to a more accurate result; see Figure~\ref{fig:kodiakHKMT}.

\subsection{Uniform $\mathrm{P}$: Linear Spreading Speed and Average Arrival Times}

The heat kernel estimate gives a prediction for each pair of cities, $n$ and $m$.  In contrast, the linear spreading speed prediction gives one prediction for the entire network.  In this section, we will compare the linear spreading speed to the average speed of infection in the network as follows.  We first separate cities into groups according their distance from the initial infection location.  We then compute the mean arrival time at each distance and compute a linear regression line using these data points.  We exclude both the original infection location and all cities in Papa New Guinea in these calculation.  For this version of the WAN, Papa New Guinea contains a sequence of cities that are the furthest distance from a majority of cities in the network; see those cities in Figure~\ref{fig:kodiakLSS} with distance nine and greater from the origin city.  Including them in the averaged arrival times gives undue importance to these nodes, so we exclude them here.  The reciprocal of the slope of the regression line then gives an estimate for the average time required for the disease to spread between adjacent nodes on the graph.  We claim that $\frac{1}{s_{lin}}$ provides a reasonable  leading order approximation.

\begin{figure}[ht]
\centering
\includegraphics[width=0.4\textwidth]{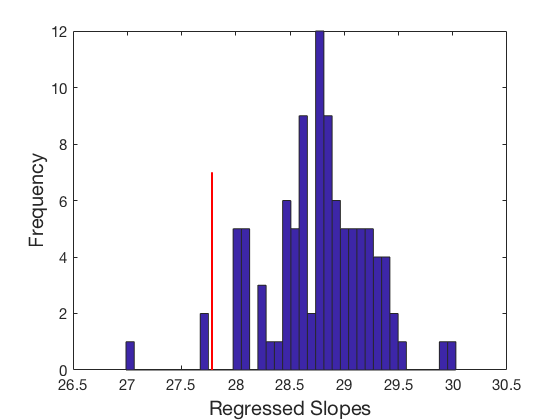} 
\includegraphics[width=0.4
\textwidth]{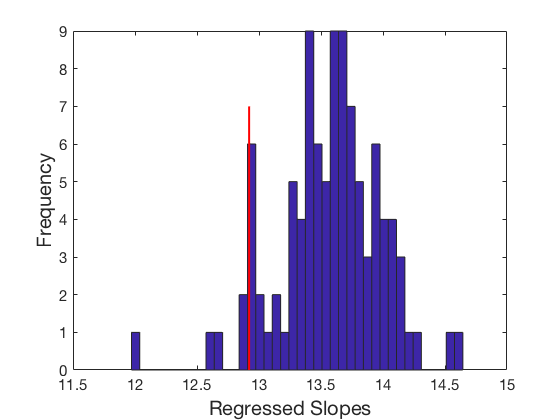} \hfil
\caption{Histogram of regressed slopes taken from 100 random cities from using uniform $\mathrm{P}=\mathrm{D}^{-1}\mathrm{A}$. The upper panel is $\gamma=0.0001$ and the lower is $\gamma=0.01$. The red line denotes the prediction based upon the linear spreading speed.  }\label{fig:uniformLSShist} 
\end{figure}

In Figure~\ref{fig:uniformLSShist}, we compare averaged inter-city invasion times to the estimate from the linear spreading speed. Using the uniform $\mathrm{P}$ matrix, we randomly select one hundred cities in which to initialize the disease and compute average invasion times.  Comparing two different values of $\gamma$, we find that the linear spreading speed is a reasonable leading order estimate. We also compile comparisons between arrival times observed  in numerical simulation to linear spreading speed for infection starting in Paris and Kodiak over varying $\gamma$ together with relative errors (REs) in Table I. All units are in days. Again, regression times were calculated disregarding the origin city and the cities in Papa New Guinea. As expected, the average arrival times are better approximations for smaller values of $\gamma$.  

\begin{table}
 \begin{tabular}{||c c c c c c||} 
 \hline
 $\gamma$ & LSS & Kodiak  & RE Kodiak & Paris & RE Paris \\ [0.5ex] 
 \hline\hline
 0.001& 19.961 & 20.779& 0.0394 & 19.223 &0.0383  \\ 
 \hline
 0.01 & 12.920  & 13.435& 0.0383 & 11.950 &0.0811  \\
 \hline
 0.1 & 8.036 & 7.636 &  0.0522& 6.281 &0.2791  \\ [1ex] 
 \hline
\end{tabular}\label{tbl:LSS1}
\caption{Arrival time per flight calculated from the linear spreading speed and from numerical observations for a range of $\gamma$ values and epidemics originating in Kodiak and Alaska.  }
\end{table}

\subsection{Proportional and Random $\mathrm{P}$}
A major challenge in estimating arrival times, even when $\gamma$ is very small, is attaining precise information regarding the weights in the matrix $\mathrm{P}$.  It is often the case that the edges of the network can be determined, but that estimating fluxes and the corresponding weights is more difficult. 

\begin{figure}[ht]
\centering
\includegraphics[width=0.3\textwidth]{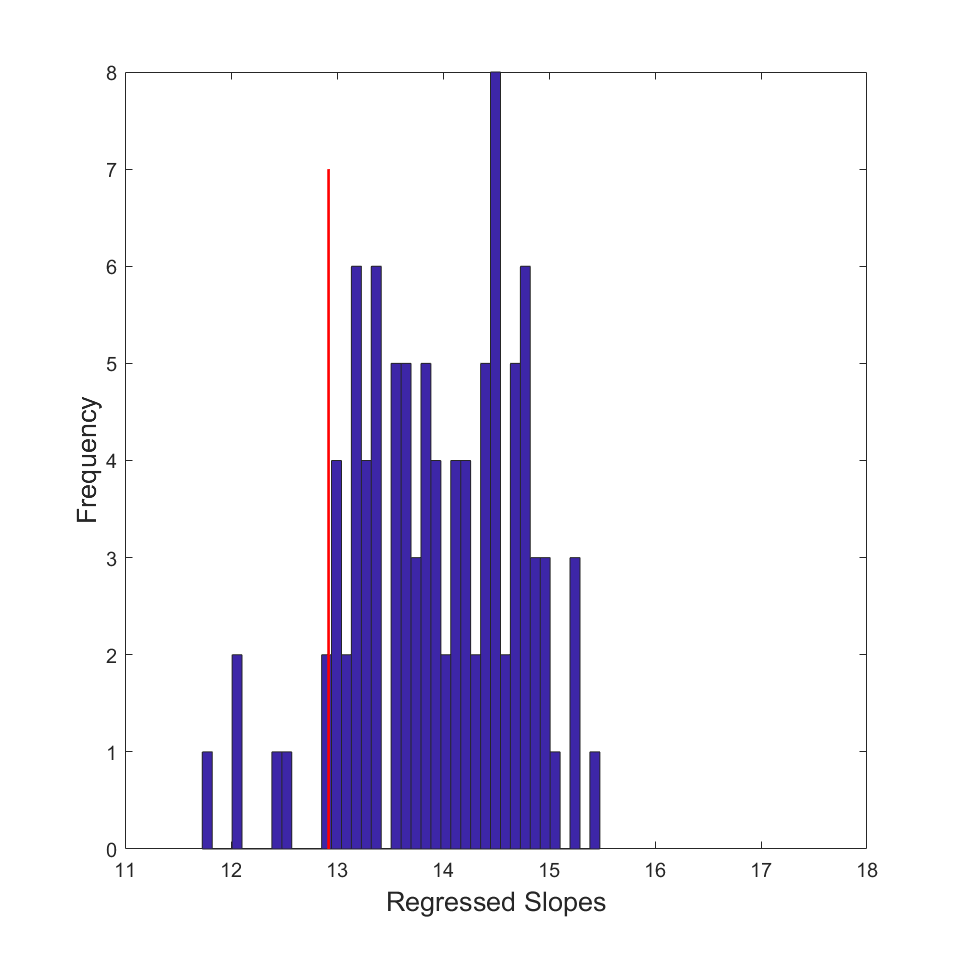} 
\includegraphics[width=0.3\textwidth]{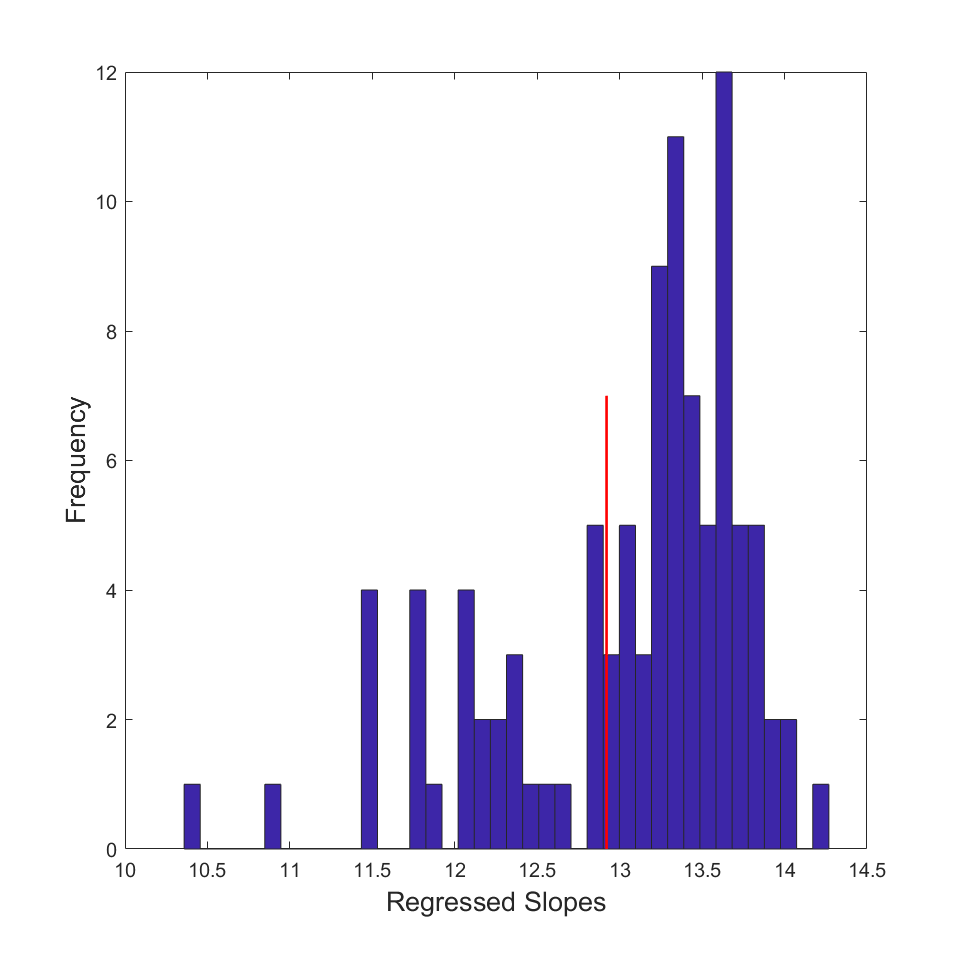} \hfil
\caption{Histogram of regressed slopes taken from 100 random cities with weights in $\mathrm{P}$ chosen randomly (upper) and proportionally (lower).  In both cases, $\gamma=0.01$. The red line denotes the value predicted by the linear spreading speed}\label{fig:randpropLSShist} 
\end{figure}

In light of the success of the heat kernel in estimating arrival times for small $\gamma$, we must comment on the utility of the linear spreading speed estimate.  A major advantage of the linear spreading speed estimate is that it does not require information regarding the topology or weights of a network, but only knowledge of the distance of the shortest paths connecting any two nodes.  In Figure~\ref{fig:randpropLSShist} we compile average invasion times found via linear regression and compare with the linear spreading speed.  In both cases, the linear spreading speed provides a reasonable leading order approximation without any knowledge of the weights in the network aside from the average degree.

\begin{figure}[ht]
\centering
\includegraphics[width=0.4\textwidth]{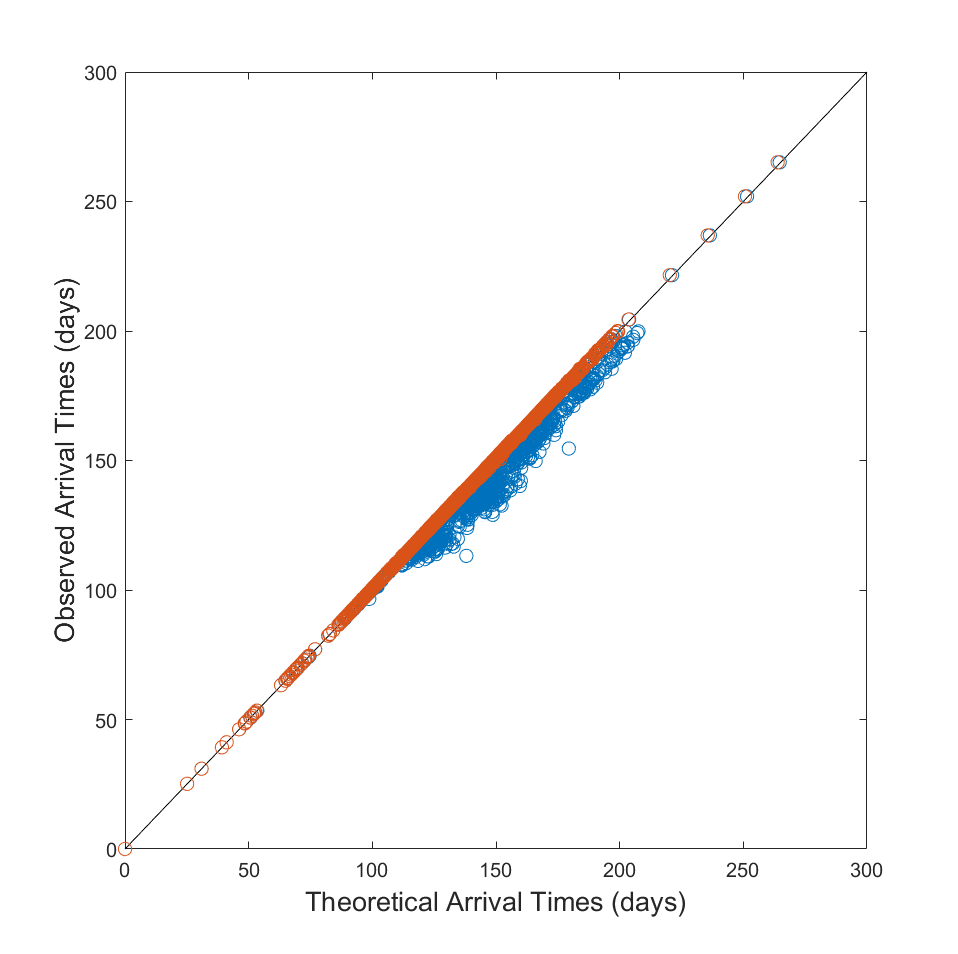} \hfil
\caption{Simulated arrival times compared to one and two-term heat kernel approximations plotted on the line $y=x$ for an infection started in Boriziny, Madagascar with $\gamma=0.001$. One term approximation is shown in blue and two is shown in orange.}\label{fig:randomPproblems} 
\end{figure}

If one does have an estimate for the weights of the network, then the heat kernel method can once again be employed to estimate arrival times.  We comment briefly on the dynamics for proportional and random weights.  Generally speaking, arrival times for proportional $\mathrm{P}$ are more accurate.  This makes intuitive sense, as most pairs of nodes have shortest paths connecting them through high degree nodes and proportional weights pushes a higher proportion of travelers along these routes.  In contrast, random $\mathrm{P}$ matrices can lead to large errors even for small values of $\gamma$, depending on the realization of the network.  We show the results of one such problematic example in Figure~\ref{fig:randomPproblems}.  Here we have $\gamma=0.001$.   The initial infection city there is Boriziny, Madagascar, chosen due to the large deviations between predicted and observed arrival times.  The reason for the deviations in this case are easy to locate.  There are only two connections to Boriziny and the random weights selected in this realization give much larger weight to one of them.  Therefore any city whose shortest path to Boriziny connects through the city with less weight will exhibit errors.  However, since the two airports connecting to Boriziny are themselves connected by a flight means that the most probable path requires one more flight than
the shortest path.  Including one more term in the heat kernel approximation leads to a dramatic increase in the accuracy of the approximation, see Figure~\ref{fig:randomPproblems}.

\section{Conclusions}\label{sec:conclusion}
We have presented two estimates for arrival times of an infectious disease in a SIR meta-population model defined on a global airline network.  Both estimates are valid in the limit as the rates of diffusion are dominated by the rates of reaction.  The linear spreading speed estimate provides a single number for the entire network and can be used to determine the leading order dependence of the arrival times on system parameters.  The heat kernel estimate requires more information regarding the network structure and provides a formula, in term of the Lambert-W function, for arrival times between any two nodes in the network.

In the direction of more realistic models of epidemics it would be interesting to understand the role of heterogeneities in system parameters, stochastic effects, time varying connections and finite size effects.  For example, city dependent reaction and recovery rates $\alpha_n$, $\beta_n$ could likely be incorporated into the heat kernel estimate in a rather straightforward way.  There has also been recent interest in determining to what extend diffusion is a suitable model for passenger flux; see for example \cite{belik11}.  

Our methods are not limited to the model considered here, but rely on the linear determinacy of the system.  This is a property shared by many systems and the methods here could likely be employed to determine arrival times in other systems and for other networks.  We mention  the Fisher-KPP equation, which combines local logistic growth with diffusion in a spatially extended environment; see for example \cite{burioni12,hoffman17}.  At the same time,  there are many examples of systems which are not linearly determinate and for which the methods developed here would not directly apply.  It would be interesting to see whether these methods could be adapted to those classes of problems.

\section*{Acknowledgments}
 This research was conducted as part of a NSF sponsored REU that was held at George Mason University during the summer of 2017.  All authors received support from the NSF (DMS-1516155).

\bibliographystyle{abbrv}
\bibliography{SIR_Arrival_Times}

\begin{thebibliography}{10}

\bibitem{barrat08}
A.~Barrat, M.~Barth\'elemy, and A.~Vespignani.
\newblock {\em Dynamical processes on complex networks}.
\newblock Cambridge University Press, Cambridge, 2008.

\bibitem{belik11}
V.~Belik, T.~Geisel, and D.~Brockmann.
\newblock Natural human mobility patterns and spatial spread of infectious
  diseases.
\newblock 1, 03 2011.

\bibitem{Brockmann2013}
D.~Brockmann and D.~Helbing.
\newblock The hidden geometry of complex, network-driven contagion phenomena.
\newblock {\em Science}, 342(6164):1337--1342, 2013.

\bibitem{burioni12}
R.~Burioni, S.~Chibbaro, D.~Vergni, and A.~Vulpiani.
\newblock Reaction spreading on graphs.
\newblock {\em Phys. Rev. E}, 86:055101, Nov 2012.

\bibitem{chen17}
Y.-Y. Chen, J.-S. Guo, and F.~Hamel.
\newblock Traveling waves for a lattice dynamical system arising in a diffusive
  endemic model.
\newblock {\em Nonlinearity}, 30(6):2334, 2017.

\bibitem{Colizza2007}
V.~Colizza, A.~Barrat, M.~Barth{\'e}lemy, and A.~Vespignani.
\newblock Predictability and epidemic pathways in global outbreaks of
  infectious diseases: the sars case study.
\newblock {\em BMC Medicine}, 5(1):34, Nov 2007.

\bibitem{Colizza06}
V.~Colizza, A.~Barrat, M.~Barthélemy, and A.~Vespignani.
\newblock The role of the airline transportation network in the prediction and
  predictability of global epidemics.
\newblock {\em Proceedings of the National Academy of Sciences of the United
  States of America}, 103(7):2015--2020, 2006.

\bibitem{Crepey2006}
P.~Cr\'epey, F.~P. Alvarez, and M.~Barth\'elemy.
\newblock Epidemic variability in complex networks.
\newblock {\em Phys. Rev. E}, 73:046131, Apr 2006.

\bibitem{diekmann00}
O.~Diekmann and J.~A.~P. Heesterbeek.
\newblock {\em Mathematical Epidemiology of Infectious Diseases: Model
  Building, Analysis and Interpretation}.
\newblock Wiley, 1 edition, 2000.

\bibitem{Gautreau2007}
A.~Gautreau, A.~Barrat, and M.~Barth{\'e}lemy.
\newblock Arrival time statistics in global disease spread.
\newblock {\em Journal of Statistical Mechanics: Theory and Experiment},
  2007(09):L09001, 2007.

\bibitem{Gautreau2008}
A.~Gautreau, A.~Barrat, and M.~Barth{\'e}lemy.
\newblock Global disease spread: Statistics and estimation of arrival times.
\newblock {\em Journal of Theoretical Biology}, 251(3):509 -- 522, 2008.

\bibitem{grais03}
R.~F. Grais, J.~Hugh~Ellis, and G.~E. Glass.
\newblock Assessing the impact of airline travel on the geographic spread of
  pandemic influenza.
\newblock {\em European Journal of Epidemiology}, 18(11):1065--1072, Nov 2003.

\bibitem{Guimera2005}
R.~{Guimer{\`a}}, S.~{Mossa}, A.~{Turtschi}, and L.~A.~N. {Amaral}.
\newblock {From the Cover: The worldwide air transportation network: Anomalous
  centrality, community structure, and cities' global roles}.
\newblock {\em Proceedings of the National Academy of Science}, 102:7794--7799,
  May 2005.

\bibitem{hindes13}
J.~Hindes, S.~Singh, C.~R. Myers, and D.~J. Schneider.
\newblock Epidemic fronts in complex networks with metapopulation structure.
\newblock {\em Phys. Rev. E}, 88:012809, Jul 2013.

\bibitem{hoffman17}
A.~Hoffman and M.~Holzer.
\newblock Invasion fronts on graphs: the {F}isher-{KPP} equation on homogeneous
  trees and {E}rdos-{R}\'{e}yni graphs.
\newblock {\em preprint}, 2017.

\bibitem{holzer14}
M.~Holzer and A.~Scheel.
\newblock Criteria for pointwise growth and their role in invasion processes.
\newblock {\em J. Nonlinear Sci.}, 24(4):661--709, 2014.

\bibitem{hosono95}
Y.~Hosono and B.~Ilyas.
\newblock Traveling waves for a simple diffusive epidemic model.
\newblock {\em Math. Models Methods Appl. Sci.}, 5(7):935--966, 1995.

\bibitem{hufnagel04}
L.~Hufnagel, D.~Brockmann, and T.~Geisel.
\newblock Forecast and control of epidemics in a globalized world.
\newblock {\em Proceedings of the National Academy of Sciences of the United
  States of America}, 101(42):15124--15129, 2004.

\bibitem{Ianelli2017}
F.~Iannelli, A.~Koher, D.~Brockmann, P.~H\"ovel, and I.~M. Sokolov.
\newblock Effective distances for epidemics spreading on complex networks.
\newblock {\em Phys. Rev. E}, 95:012313, Jan 2017.

\bibitem{kermack32}
W.~O. Kermack and A.~G. McKendrick.
\newblock Contributions to the mathematical theory of epidemics. ii. the
  problem of endemicity.
\newblock {\em Proceedings of the Royal Society of London A: Mathematical,
  Physical and Engineering Sciences}, 138(834):55--83, 1932.

\bibitem{newman03}
M.~E.~J. Newman.
\newblock The structure and function of complex networks.
\newblock {\em SIAM Review}, 45(2):167--256, 2003.

\bibitem{pastor15}
R.~Pastor-Satorras, C.~Castellano, P.~Van~Mieghem, and A.~Vespignani.
\newblock Epidemic processes in complex networks.
\newblock {\em Rev. Mod. Phys.}, 87:925--979, Aug 2015.

\bibitem{porter16}
M.~A. Porter and J.~P. Gleeson.
\newblock {\em Dynamical systems on networks}, volume~4 of {\em Frontiers in
  Applied Dynamical Systems: Reviews and Tutorials}.
\newblock Springer, Cham, 2016.
\newblock A tutorial.

\bibitem{rvachev85}
L.~A. Rvachev and I.~M. Longini.
\newblock A mathematical model for the global spread of influenza.
\newblock {\em Mathematical Biosciences}, 75(1):3 -- 22, 1985.

\bibitem{vansaarloos03}
W.~van Saarloos.
\newblock Front propagation into unstable states.
\newblock {\em Physics Reports}, 386(2-6):29 -- 222, 2003.

\bibitem{Wu2006}
Z.~Wu, L.~A. Braunstein, V.~Colizza, R.~Cohen, S.~Havlin, and H.~E. Stanley.
\newblock Optimal paths in complex networks with correlated weights: The
  worldwide airport network.
\newblock {\em Phys. Rev. E}, 74:056104, Nov 2006.

\end{thebibliography}

\end{document}